# On undesired emergent behaviors in compound prostate cancer detection systems

Erlend Sortland Rolfsnes[1], Philip Thangngat[1], Trygve Eftestøl[1], Tobias Nordström[2,3], Fredrik Jäderling[4,5], Martin Eklund[2], and Alvaro Fernandez-Quilez[1,2]

[1] Department of Electrical Engineering and Computer Science, University of Stavanger, Norway.
[2] Department of Medical Epidemiology and Biostatistics, Karolinska Institutet, Sweden.
[3] Department of Clinical Sciences, Danderyd Hospital, Sweden.
[4] Department of Radiology, Capio Saint Göran Hospital, Sweden.
[5] Department of Molecular Medicine and Surgery, Karolinska Institutet, Sweden.
alvaro.f.quilez@uis.no

**Abstract.** Artificial intelligence systems show promise to aid in the diagnostic pathway of prostate cancer (PC), by supporting radiologists in interpreting magnetic resonance images (MRI) of the prostate. Most MRI-based systems are designed to detect clinically significant PC lesions, with the main objective of preventing over-diagnosis. Typically, these systems involve an automatic prostate segmentation component and a clinically significant PC lesion detection component. In spite of the compound nature of the systems, evaluations are presented assuming a standalone clinically significant PC detection component. That is, they are evaluated in an idealized scenario and under the assumption that a highly accurate prostate segmentation is available at test time. In this work, we aim to evaluate a clinically significant PC lesion detection system accounting for its compound nature. For that purpose, we simulate a realistic deployment scenario and evaluate the effect of two non-ideal and previously validated prostate segmentation modules on the PC detection ability of the compound system. Following, we compare them with an idealized setting, where prostate segmentations are assumed to have no faults. We observe significant differences in the detection ability of the compound system in a realistic scenario and in the presence of the highest-performing prostate segmentation module (DSC: 90.07±0.74), when compared to the idealized one (AUC: 77.97±3.06 and 84.30±4.07, $P<.001$). Our results depict the relevance of holistic evaluations for PC detection compound systems, where interactions between system components can lead to decreased performance and degradation at deployment time.

## 1    Introduction

Over the last decade, magnetic resonance imaging (MRI) has become an important tool for prostate cancer (PC) detection, staging and treatment planning [4]. Its surge in relevance for PC diagnosis is expected to substantially increase the amount of imaging examinations, and, consequently, the radiologist workload [7, 11]. The increasing volume of MRI coupled with a shortage of specialists, presents a favorable scenario to delegate time-consuming tasks such as the detection of clinically significant PC (csPC) lesions to artificial intelligence (AI) systems [18].

Whilst AI systems for csPC detection hold potential to augment the radiologist abilities, reduce reader variability and shorten the study times, their availability is not translated into clinical adoption [1,24]. Limitations in adopting these systems are often reported to stem from a misalignment between system evaluation and its intended use in clinical practice [12]. This misalignment is typically exemplified by AI systems exhibiting an overoptimistic performance during their development that does not match their performance at deployment time [10, 23, 24].

AI systems for csPC detection are typically compound systems that integrate different interacting components [18]. In particular, most csPC detection systems rely on an AI prostate gland segmentation component followed by a csPC detection module that depends on the quality of the previously segmented prostate [22, 23]. In spite of the dependence and expected interaction of the components, the evaluation of the compound system is performed assuming a standalone detection module. That is, the compound system is evaluated in an idealized scenario and under the assumption of the availability of a highly accurate prostate segmentation [22]. Hence, without accounting for the emergent behaviors due to the components' interaction and resulting in an evaluation that is not representative of the system as a whole.

Against this background, we aimed at performing a holistic evaluation of a typical compound csPC detection system that integrates a prostate segmentation module and a detection module [8, 18]. Further, we aim to investigate emergent behaviors that stem from unexpected results that propagate from the prostate segmentation module to the csPC detection module (Figure 1). We hypothesized that unaccounted emergent behaviors stemming from the interaction between modules might be associated with an overoptimistic report in the performance of the system.

## 2    Related work

Multiple works have attempted to develop csPC detection compound systems. In the work by Cao *et al.* [2], the authors propose a compound system based on a 2D network for csPC detection and lesion grade prediction, reaching a 0.81 area under the curve (AUC) detection ability for MRI slices that contained lesions. In the work of Sanyal *et al.* [21], the authors proposed a compound system where the first component segments the prostate gland, whilst the second component performs csPC detection based on the previously segmented prostate (AUC = 0.86). More recently, Sanford *et al.* [20], developed a system that relies on previously segmented prostates to detect csPC from MRI. In the work, they compare the performance of the system with expert radiologists reaching a 56% agreement. Fernandez-Quilez *et al.* [8] proposed a compound system that detects PC by obtaining a surrogate biomarker of age from MRI (AUC = 0.981). In this work, the csPC detection module assumes the availability of high quality prostate segmentation masks. Finally, Saha *et al.* [18] present a compound csPC detection system based on a prostate segmentation module and an ensemble of different architectures based on the largest data sample to date for csPC detection (AUC = 0.91). The system outperformed a consensus of 62 radiologist specialists. Although these methods achieve a high performance, they all share a lack of evaluation in terms of interaction between the modules that compose the proposed compound systems.

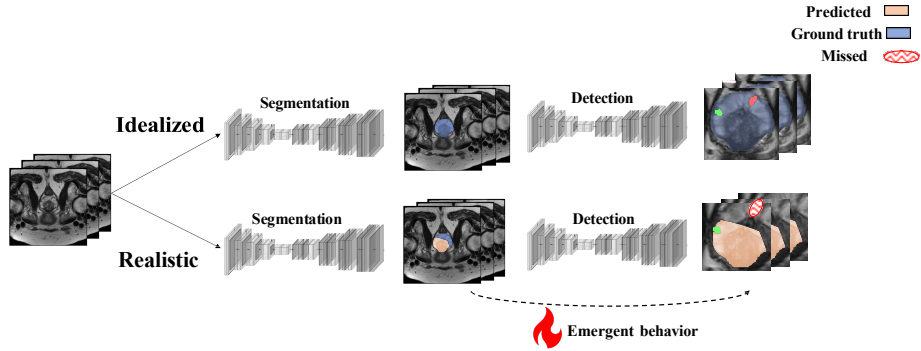

**Fig. 1.** Compound clinically-significant prostate cancer detection system evaluated under the assumption that no faults will be present in the output of the segmentation module (idealized), and accounting for the behaviors emerging from the interaction of the segmentation and detection modules (realistic).

## 3  Materials and Methods

### 3.1  Study cohort

In our retrospective single-center study, we leverage data from the open access ProstateX challenge data (Radboud University, Netherlands) [15]. ProstateX consists of a collection of prostate MRI exams collected with the purpose of validating modern AI algorithms for the diagnostic classification of csPC. Subjects were recruited on the basis of suspicion of PC based on high PSA levels. All subjects had PC diagnosis confirmed through an MRI-guided biopsy [14].

The ProstateX challenge cohort consisted of 204 subjects (median age 66 years [range, 48-83]), with available T2-weighted (T2w) spin sequences and diffusion-weighted (DW) magnetic resonance exams. The cohort had pixel-level annotations for the prostate gland and prostate lesions obtained by two experienced board-certified radiologists with > 5 years of experience [3]. All annotations were obtained with ITK-SNAP v.380 software (http://www.itksnap.org/) [25].

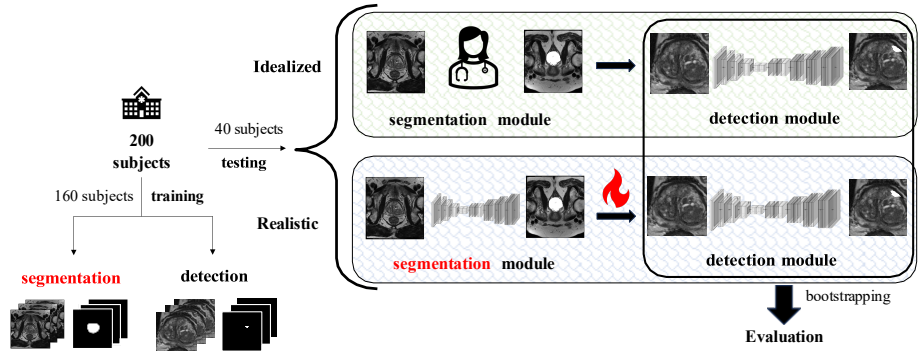

**Fig. 2.** Technical development of the project, depicting the splitting of the data, MRI exams and ground truth used to train the segmentation and detection module and their test protocol. Whilst in the idealized case it is assumed that the segmentation is absent of faults, in the realistic case we explore emergent behaviors stemming from segmentation faults.

**Magnetic resonance Imaging.** From the original 204 available subjects, we leverage 200 subjects 3.0 Tesla (T) axial T2w spin sequences based on the availability of both pixel-level lesion annotations and prostate gland annotations (Figure 2). Images were acquired with either Siemens (Siemens Health Engineers, Erlangen, Germany) or Philips (Philips & Co, Eindhoven, The Netherlands) scanners. The T2w MRI exams had an in-plane resolution of [0.5-0.562mm]x[0.5-0.562mm]x[3.0-3.15mm].

*Preprocessing and splitting.* To have a common space of reference, we apply a re-sampling by linear interpolation to all the T2w sequences to a resolution of 0.5x0.5x3.6mm. In addition, we also apply N4 bias field correction filtering and normalization of the T2w exams pixel intensity [13]. Following, we split the cohort in 80/20% whilst avoiding cross-contamination. The splitting resulted in 160 and 40 patients for the train and test set, respectively. The splitting is performed stratifying by PC lesion type, which results in 498 (72.80%) non-clinically significant PC (non-csPC) lesions in the train set, 186 (27.20%) in the test set, 222 (71.61%) csPC lesions in the train set and 88 (28.39%) in the test set.

### 3.2 Compound PC detection system

As shown in Figure 2, the compound system integrates two modules: automatic prostate segmentation [9, 13] and csPC lesion detection [8, 18]. We train and evaluate both modules with the same data split, to avoid undesired data leakage.

**Prostate segmentation module.** We leverage two different segmentation networks. In the first case, we focus on a multi-view segmentation network with a good generalization ability based on previously reported results [13]. In addition, we employ a 2D nnU-Net based on its wide adoption and positive results in previous prostate segmentation challenges [9]. The training follows the steps described in the original article for the multi-view network and the default configuration for the 2D nnU-Net, respectively [9, 13]. Given the objectives of this work, exploration of architectural or training modifications is considered out of the scope of the work. Training and evaluation is performed on an NVIDIA A100 GPU (NVIDIA Corporation, Santa Clara, USA).

**Clinically significant PC detection module.** Based on previously reported results [18], we adopt a standard U-Net architecture [17]. We use an ImageNet pre-trained model and a channel-wise convolutional block to map the original gray-scale images (1 channel) to the expected Red-Green-Blue (RGB) format (3 channels). Following standard approaches, we train the architecture using cropped 2D T2w slices as an input. We crop the slices by leveraging the prostate segmentation mask obtained from the previous module. The cropping is performed around the prostate gland with a 20 pixel margin to avoid edge cases and ensure that we capture the whole region of interest. We train using a 5-fold cross-validation (CV), resulting in 5 models. These models were then ensembled with equal weighting into a single system. Training and evaluation is performed on an NVIDIA A100 GPU (NVIDIA Corporation, Santa Clara, USA).

### 3.3   Emergent behaviors

Following standard practices for compound PC systems, we start by providing a standalone evaluation of the csPC detection module at test time. That is, the cropped T2w input of the csPC detection module is obtained by leveraging the prostate segmentation ground truth. This is referred as idealized scenario, where the interaction between modules is disregarded. Following, we evaluate the modules accounting for possible emergent behaviors due to their interaction. This is referred as realistic scenario. In this case, we start by characterizing the performance of the segmentation module. Next, we characterize the performance of the csPC detection module by leveraging the output of the segmentation module as input for the csPC detection module. The realistic evaluation process is repeated for the multi-view and nnU-Net architectures, with heterogeneous segmentation performances (Figure 1).

    In both scenarios and for both modules, the results are presented at the patient level. That is, the results are presented for the whole patient exam (3D). For the segmentation module, the results are presented in terms of Dice Score Coefficient (DSC, %), Relative Volume Difference (RVD, %) and Hausdorff Distance (HD, mm). For the csPC detection module, the results are presented in terms of DSC and AUC. For the AUC, lesions that shared a minimum DSC of 0.10 when compared with the ground truth were considered true positives. We follow similar criteria to the one presented in previous works [16, 19].

### 3.4   Statistical testing

We obtain an estimate of the test results metrics for the compound system using a non-parametric bootstrapping with $n = 1000$ replicates without repetition. We assess statistical significance for AUC through permutation tests. A $P$ <0.05 is considered statistically significant. All results are presented as mean ± standard deviation (SD). All analyses were performed in Python 3 (www.python.org/downloads) with the open-sourced statsmodels 0.14.0 module (www.statsmodels.org).

**Table 1.** Results for the segmentation module when independently evaluated. Results are presented for two different segmentation architectures.

| Architecture | DSC (%) ↑ | RVD (%) ↓ | HD (mm) ↓ |
|---|---|---|---|
| nnU-Net [9] | 85.98 ± 6.18 | 3.73 ± 9.02 | 7.76 ± 4.82 |
| tU-Net [13] | 90.07 ± 0.74 | 2.01 ± 1.30 | 1.76 ± 0.39 |

↑ larger better ↓ lower better,  DSC = Dice Score Coefficient,
RVD = Relative Volume Difference, HD = Hausdorff distance.

## 4   Results and Discussion

Table 1 depicts the experimental results for the segmentation module. We can observe that nnU-Net presents a DSC that is close to expert level radiologist annotation accuracy (DSC>=0.86) [5], whilst tU-Net presents a DSC that is above that level. In terms of RVD and HD, we can observe that tU-Net presents a higher accuracy in volumetric estimation and a lower HD.

Table 2 presents the results for the compound system in an idealized and realistic scenarios, respectively. When analysing the results of the detection module as a standalone module in the idealized scenario, the compound system presents a bootstrapped AUC = 84.30 ± 4.07 for csPC lesions. When evaluated accounting for emergent behaviors stemming from the interaction between the segmentation module and the detection module, we can observe a significant decrease in the AUC performance. Specifically, when leveraging nnU-Net AUC = 71.30 ± 3.83 and when leveraging tU-Net AUC = 77.97 ± 3.06. The results show that a higher individual performance in the first module (tU-Net) is translated into a higher compound system performance.

As depicted qualitatively in Figure 3, we can observe from an intuitive perspective the effect of a faulty segmentation for the input of the detection module. In particular, we can observe the significant effect in the input of the detection module of the different segmentation module output architectures.

Our retrospective study provides some degree of evidence of the emergence of undesired behaviors when accounting for the interaction of modules in a typical compound PC detection system. In particular, for a compound system that

**Table 2.** Results for the compound system when evaluated in an idealized setting and in a realistic setting. In the first case, the detection module is evaluated as a standalone module, without accounting for faulty segmentations. In the second case, emergent behaviors are taken into account by using the output of the segmentation module as the input for the detection one.

| *Scenario* | Architecture | Lesion | DSC (%) ↑ | AUC (%) ↑ | $P_{AUC}$ |
|---|---|---|---|---|---|
| Idealized | ✗ | csPC | 12.40 ± 1.96 | 84.30 ± 4.07 | ✗ |
| | | non-csPC | 16.97 ± 1.34 | 86.90 ± 2.23 | ✗ |
| Realistic | nnU-Net [9] | csPC | 7.14 ± 2.05 | 71.30 ± 3.83 | <.001$^†$ |
| | | non-csPC | 9.84 ± 0.98 | 75.57 ± 3.61 | <.001$^†$ |
| | tU-Net [13] | csPC | 9.24 ± 3.01 | 77.97 ± 3.06 | <.001$^†$ |
| | | non-csPC | 11.08 ± 1.85 | 80.12 ± 2.71 | <.001$^†$ |

↑ larger better   DSC = Dice Score Coefficient, AUC = Area under the curve.

$†$ Reference is idealized scenario. P<0.05 were considered statistically significant.

leverages a segmentation module and a csPC detection module based on U-Net variants [13, 18, 21]. Our results support the hypothesis that unaccounted interactions are correlated with an overoptimistic report in the compound system performance.

Previous studies in the literature have tackled the csPC detection from MRI with compound systems, yielding to AUCs in the range of 0.81-0.91 [2, 18, 21]. However, the evaluation of the system is presented in terms of the detection ability as a standalone module and without accounting for faults stemming in the preceding segmentation modules. In contract to those studies, we aim at characterizing emergent behaviors stemming from the interaction of the modules in the compound PC system. Despite not being directly comparable, our results suggest significant drops in csPC detection performance when accounting for the interaction between modules.

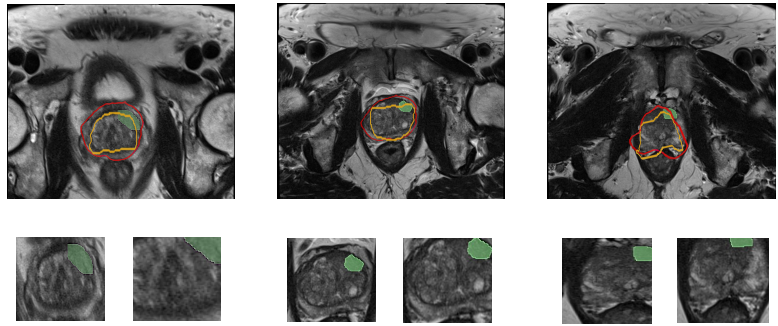

**Fig. 3.** Qualitative results from the segmentation module and in the presence of two different segmentation architectures. In yellow, results for nnU-Net. In red, results for tU-Net. Bottom row depicts the eventual input for the detection module, based on the output of the segmentation one.

With the expected increase in AI developments based on MRI for PC [8], our study highlights the importance of a holistic evaluation for compound PC systems. In that regard, our results suggest that the lack of exhaustive evaluations can lead to overoptimistic evaluation reports. Hereby, a comprehensive independent evaluation of each module and of the interaction between modules could benefit the translation of the systems into clinical practice [6, 24].

**Limitations.** First, our study was limited by its retrospective nature and lack of external evaluation. Second, our study was limited to compound PC systems consisting of a segmentation module and a detection module. In spite of their wide adoption, other configurations should be considered in future studies. Finally, we restricted our study to U-Net based architectures. Future studies considering other widely adopted architectures for PC detection might provide additional insights about the interactions effect in compound PC systems.

## 5    Conclusion

In conclusion, our results indicate that unaccounted interactions between modules of compound csPC detection systems might lead to overoptimistic evaluation results. We found that when accounting for the interaction between modules, undesired behaviors emerge leading to significant drops in performance. To avoid it, we suggest that evaluations of PC compound systems should be holistic, presenting the performance of independent modules and when interacting. Our observations could potentially increase the adoption of AI developments for PC and MRI in clinical practice.

**Acknowledgments.** The authors would like to show their gratitude to the organizers of the ProstateX challenge for providing access to their curated dataset.

**Disclosure of Interests.** The authors declare that they have no competing interests.

## References

1. Anaya-Isaza, A., Mera-Jiménez, L., Fernandez-Quilez, A.: Crosstransunet: A new computationally inexpensive tumor segmentation model for brain mri. IEEE Access **11**, 27066–27085 (2023)

2. Cao, R., Bajgiran, A.M., Mirak, S.A., Shakeri, S., Zhong, X., Enzmann, D., Raman, S., Sung, K.: Joint prostate cancer detection and gleason score prediction in mp-mri via focalnet. IEEE transactions on medical imaging **38**(11), 2496–2506 (2019)

3. Cuocolo, R., Comelli, A., Stefano, A., Benfante, V., Dahiya, N., Stanzione, A., Castaldo, A., De Lucia, D.R., Yezzi, A., Imbriaco, M.: Deep learning whole-gland and zonal prostate segmentation on a public mri dataset. Journal of Magnetic Resonance Imaging **54**(2), 452–459 (2021)

4. Eklund, M., Jäderling, F., Discacciati, A., Bergman, M., Annerstedt, M., Aly, M., Glaessgen, A., Carlsson, S., Grönberg, H., Nordström, T.: Mri-targeted or standard biopsy in prostate cancer screening. New England Journal of Medicine **385**(10), 908–920 (2021)

5. Fassia, M.K., Balasubramanian, A., Woo, S., Vargas, H.A., Hricak, H., Konukoglu, E., Becker, A.S.: Deep learning prostate mri segmentation accuracy and robustness: A systematic review. Radiology: Artificial Intelligence p. e230138 (2024)

6. Fernandez-Quilez, A.: Deep learning in radiology: ethics of data and on the value of algorithm transparency, interpretability and explainability. AI and Ethics **3**(1), 257–265 (2023)

7. Fernandez-Quilez, A., Larsen, S.V., Goodwin, M., Gulsrud, T.O., Kjosavik, S.R., Oppedal, K.: Improving prostate whole gland segmentation in t2-weighted mri with synthetically generated data. In: 2021 IEEE 18th International Symposium on Biomedical Imaging (ISBI). pp. 1915–1919. IEEE (2021)

8. Fernandez-Quilez, A., Nordström, T., Jäderling, F., Kjosavik, S.R., Eklund, M.: Prostate age gap: An mri surrogate marker of aging for prostate cancer detection. Journal of Magnetic Resonance Imaging **60**(2), 458–468 (2024). https://doi.org/10.1002/jmri.29090

9. Isensee, F., Jaeger, P.F., Kohl, S.A., Petersen, J., Maier-Hein, K.H.: nnu-net: a self-configuring method for deep learning-based biomedical image segmentation. Nature methods **18**(2), 203–211 (2021)

10. Kurbatskaya, A., Jaramillo-Jimenez, A., Ochoa-Gomez, J.F., Brønnick, K., Fernandez-Quilez, A.: Assessing gender fairness in eeg-based machine learning detection of parkinson's disease: A multi-center study. In: 2023 31st European Signal Processing Conference (EUSIPCO). pp. 1020–1024 (2023). https://doi.org/10.23919/EUSIPCO58844.2023.10289837

11. Kwee, T.C., Kwee, R.M.: Workload of diagnostic radiologists in the foreseeable future based on recent scientific advances: growth expectations and role of artificial intelligence. Insights into imaging **12**(1), 1–12 (2021)

12. van Leeuwen, K.G., Schalekamp, S., Rutten, M.J., van Ginneken, B., de Rooij, M.: Artificial intelligence in radiology: 100 commercially available products and their scientific evidence. European radiology **31**, 3797–3804 (2021)

13. Lindeijer, T.N., Ytredal, T.M., Eftestøl, T., Nordström, T., Jäderling, F., Eklund, M., Fernandez-Quilez, A.: Leveraging multi-view data without annotations for prostate mri segmentation: A contrastive approach. arXiv preprint arXiv:2308.06477 (2023)

14. Litjens, G., Debats, O., Barentsz, J., Karssemeijer, N., Huisman, H.: Computer-aided detection of prostate cancer in mri. IEEE transactions on medical imaging **33**(5), 1083–1092 (2014)

15. Litjens, G., Toth, R., Van De Ven, W., Hoeks, C., Kerkstra, S., Van Ginneken, B., Vincent, G., Guillard, G., Birbeck, N., Zhang, J., et al.: Evaluation of prostate segmentation algorithms for mri: the promise12 challenge. Medical image analysis **18**(2), 359–373 (2014)

16. McKinney, S.M., Sieniek, M., Godbole, V., Godwin, J., Antropova, N., Ashrafian, H., Back, T., Chesus, M., Corrado, G.S., Darzi, A., et al.: International evaluation of an ai system for breast cancer screening. Nature **577**(7788), 89–94 (2020)

17. Ronneberger, O., Fischer, P., Brox, T.: U-net: Convolutional networks for biomedical image segmentation. In: Medical Image Computing and Computer-Assisted Intervention–MICCAI 2015: 18th International Conference, Munich, Germany, October 5-9, 2015, Proceedings, Part III 18. pp. 234–241. Springer (2015)

18. Saha, A., Bosma, J.S., Twilt, J.J., van Ginneken, B., Bjartell, A., Padhani, A.R., Bonekamp, D., Villeirs, G., Salomon, G., Giannarini, G., et al.: Artificial intelligence and radiologists in prostate cancer detection on mri (pi-cai): an international, paired, non-inferiority, confirmatory study. The Lancet Oncology (2024)

19. Saha, A., Hosseinzadeh, M., Huisman, H.: End-to-end prostate cancer detection in bpmri via 3d cnns: effects of attention mechanisms, clinical priori and decoupled false positive reduction. Medical image analysis **73**, 102155 (2021)

20. Sanford, T., Harmon, S.A., Turkbey, E.B., Kesani, D., Tuncer, S., Madariaga, M., Yang, C., Sackett, J., Mehralivand, S., Yan, P., et al.: Deep-learning-based artificial intelligence for pi-rads classification to assist multiparametric prostate mri interpretation: a development study. Journal of Magnetic Resonance Imaging **52**(5), 1499–1507 (2020)

21. Sanyal, J., Banerjee, I., Hahn, L., Rubin, D.: An automated two-step pipeline for aggressive prostate lesion detection from multi-parametric mr sequence. AMIA Summits on Translational Science Proceedings **2020**, 552 (2020)

22. Suarez-Ibarrola, R., Sigle, A., Eklund, M., Eberli, D., Miernik, A., Benndorf, M., Bamberg, F., Gratzke, C.: Artificial intelligence in magnetic resonance imaging–based prostate cancer diagnosis: where do we stand in 2021? European Urology Focus **8**(2), 409–417 (2022)

23. Wenderott, K., Krups, J., Luetkens, J.A., Gambashidze, N., Weigl, M.: Prospective effects of an artificial intelligence-based computer-aided detection system for prostate imaging on routine workflow and radiologists' outcomes. European Journal of Radiology **170**, 111252 (2024)

24. Widner, K., Virmani, S., Krause, J., Nayar, J., Tiwari, R., Pedersen, E.R., Jeji, D., Hammel, N., Matias, Y., Corrado, G.S., et al.: Lessons learned from translating ai from development to deployment in healthcare. Nature Medicine pp. 1–3 (2023)

25. Yushkevich, P.A., Gao, Y., Gerig, G.: Itk-snap: An interactive tool for semi-automatic segmentation of multi-modality biomedical images. In: 2016 38th annual international conference of the IEEE engineering in medicine and biology society (EMBC). pp. 3342–3345. IEEE (2016)